\documentclass[final,3p,onecolumn,12pt]{elsarticle}
\usepackage{graphicx}
\usepackage{amssymb}
\usepackage{epstopdf}
\usepackage{color}
\usepackage{lineno,hyperref}
\usepackage[colorinlistoftodos]{todonotes}
\usepackage{lscape}
\journal{***}
\begin{document}
\begin{frontmatter}
\title{Elastic Properties of Graphyne-based Nanotubes}
\author[primeiro]{J. M. De Sousa\corref{author2}}
\author[Trento1]{R. A. Bizao}
\author[endereco]{V. P. Sousa Filho}
\author[endereco]{A. L. Aguiar}
\author[mysecondaryaddress]{V. R. Coluci}
\author[Trento1,Trento3]{N. M. Pugno}
\author[endereco]{E. C. Girao}
\author[adressUFC]{A. G. Souza Filho}
\author[mymainaddress]{D. S. Galvao\corref{author1}}
\cortext[author1]{Corresponding author}
\cortext[author2]{Corresponding author}
\ead{galvao@ifi.unicamp.br}
\ead{josemoreiradesousa@ifpi.edu.br}
\address[primeiro]{Instituto Federal do  Piau\'i - IFPI, S\~ao Raimundo Nonato, Piau\'i, 64770-000, Brazil}
\address[endereco]{Departamento de F\'isica, Universidade Federal do Piau\'i, Teresina, Piau\'i, 64049-550, Brazil.}
\address[mymainaddress]{Applied Physics Department and Center of Computational Engineering and Science, University of Campinas - UNICAMP, Campinas-SP 13083-959, Brazil.}
\address[Trento1]{Department of Civil, Environmental and Mechanical Engineering, Laboratory of Bio-Inspired and Graphene Nanomechanics, University of Trento, via Mesiano, 77, 38123 Trento, Italy.}
\address[Trento3]{School of Engineering and Materials Science, Queen Mary University of London, Mile End Road, London E1 4NS, United Kingdom.}
\address[adressUFC]{Departamento de F\'ísica, Universidade Federal de Cear\'a, Fortaleza, CE, 60455-760, Brazil}
\address[mysecondaryaddress]{School of Technology, University of Campinas -- UNICAMP, Limeira, 13484-332 SP, Brazil}

\begin{abstract}
Graphyne nanotubes (GNTs) are nanostructures obtained from rolled up graphyne sheets, in the same way carbon nanotubes (CNTs) are obtained from graphene ones. Graphynes are $2D$ carbon-allotropes composed of atoms in $sp$ and $sp^{2}$ hybridized states. Similarly to conventional CNTs, GNTs can present different chiralities and electronic properties. Because of the acetylenic groups (triple bonds), GNTs exhibit large sidewall pores that influence their mechanical properties. In this work, we studied the mechanical response of GNTs under tensile stress using fully atomistic molecular dynamics simulations and density functional theory (DFT) calculations. Our results show that GNTs mechanical failure (fracture) occurs at larger strain values in comparison to corresponding CNTs, but paradoxically with smaller ultimate strength and Young's modulus values. This is a consequence of the combined effects of the existence of triple bonds and increased porosity/flexibility due to the presence of acetylenic groups.
\end{abstract}


\end{frontmatter}


\section{Introduction}

Graphene became one of the most studied structures in materials science since its first experimental realization in 2004~\cite{Novoselov666}. 
The advent of graphene created a renewed interest in the investigation of other 2D carbon-based nanostructures such as carbon nitride~\cite{thomas2008graphitic}, pentagraphene~\cite{zhang2015penta}, phagraphene~\cite{wang2015phagraphene} and the so-called graphynes~\cite{baughman1987structure}, among others.
Proposed by Baughman, Eckhardt and Kertesz in 1987, graphyne is a generic name for a family of $2D$ carbon-allotropes formed by carbon atoms in $sp$ and $sp^{2}$ hybridized states connecting benzenoid-like rings~\cite{baughman1987structure}. The possibility of creating different graphyne structures with different porosities, electronic and/or mechanical properties can be exploited in several technological applications, such as energy storage~\cite{srinivasu2012graphyne, hwang2013multilayer} and water purification \cite{lin2013mechanics,kou2014graphyne}. The recent advances in synthetic routes to some graphyne-like structures~\cite{haley2008synthesis} have attracted much attention to graphyne, since theoretical calculations have revealed interesting mechanical properties of single-layer~\cite{cranford2011mechanical, zhang2012mechanical} and multi-layer graphyne~\cite{peng2012mechanical}, as well the presence of Dirac cones~\cite{kim2012graphyne}. 

\begin{figure}[ht!]
\centering
\centerline{\includegraphics[width=10cm]{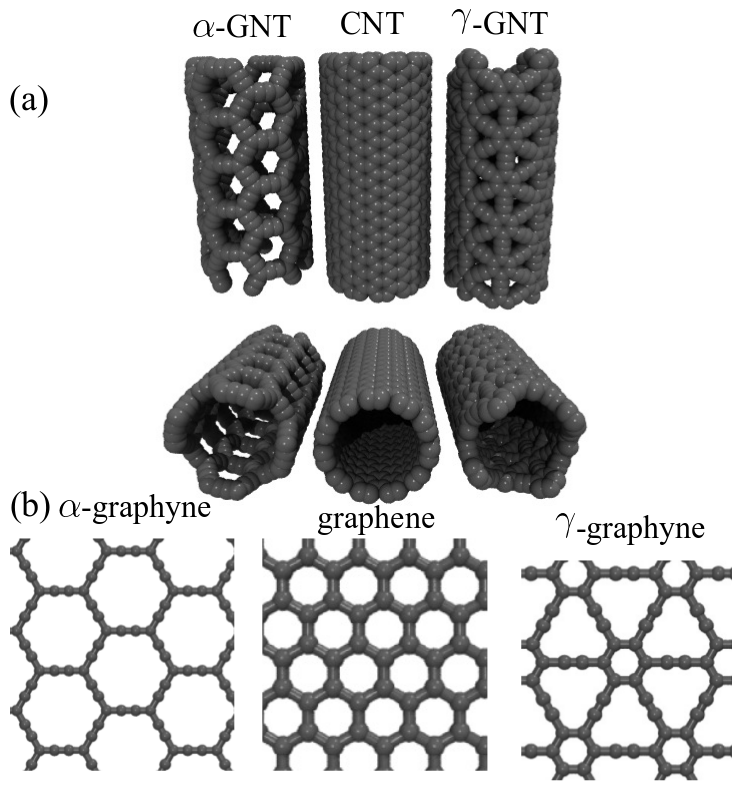}}
\caption{{\it (a) Quasi-1D nanotubes and (b) $2D$ carbon nanostructures. (a) From left to right: $\alpha$-GNT$(4,4)$, CNT$(11,11)$, and  $\gamma$-GNT$(4,4)$. (b) $2D$ carbon sheets that generated the above nanotubes.}} 
\label{figure1}
\end{figure}

Similarly to $2D$, quasi-$1D$ carbon structures have also received special attention in the last decades. For example, CNTs have been used as field-emission electron sources~\cite{de1995carbon}, tissue scaffolds~\cite{harrison2007carbon}, 
actuators~\cite{baughman1999carbon}, and artificial muscles~\cite{lima2012electrically}. Because CNTs can be conceptually 
seen as graphene sheets rolled up into cylindrical form~\cite{iijima1991helical,dresselhaus1995physics,saito1998physical}, the same concept has been used to propose graphyne-based nanotubes (GNTs). Preserving the same CNT $(n,m)$ nomenclature to describe nanotubes of different chiralities, different GNT families were theoretically predicted by Coluci \textit{et. al.}~\cite{coluci2003families,coluci2004new,coluci2004theoretical} (Fig.~\ref{figure1}). GNTs exhibit different electronic properties in comparison to CNTs, for instance, $\gamma$-GNTs are predicted to have the same band gap for any diameter \cite{li2018mechanical}. There is a renewed interest in their electronic properties \cite{Reihani2018}. Likewise the electronic behavior and mechanical properties of GNTs also show interesting features \cite{Zhao2015,Yang2012}. For example, molecular dynamics (MD) simulations have shown that, under twisting deformations, GNTs would be superplastic and more flexible than CNTs, with fracture occurring at angles three times larger than those of CNTs~\cite{de2016torsional}.
In another recent MD work \cite{li2018mechanical} carried out with AIREBO potential, the mechanical properties of graphynes-based nanotubes of $\gamma$ type ($\gamma$-GNTs) were predicted to not be very sensitive to their length and to the strain rate, while the Young's modulus ($Y$) values increase with larger diameters.

Although there is a great interest in the properties of GNTs, a fully comprehensive investigation of their mechanical properties has not been yet fully carried out and it is one the objectives of the present work. In this work we have investigated the behavior of GNTs under mechanical tensile stress using fully atomistic reactive molecular dynamics (MD) and density functional theory (DFT) simulations.

\section{Methodology}

\subsection{Molecular Dynamics Simulations}

\begin{figure}
\centering
\centerline{\includegraphics[width=14cm]{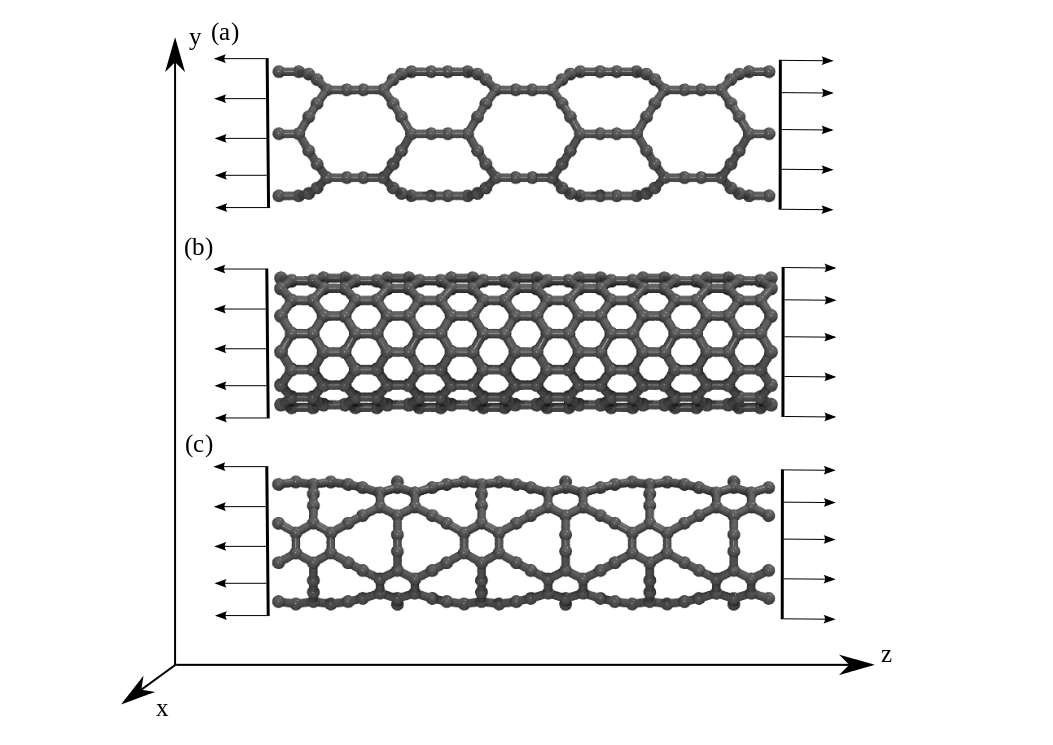}}
\caption{{\it Schematic of the tensile stress/strain simulations for: (a) $\alpha$-GNT; (b) CNT, and; (c) $\gamma$-GNT. The arrows indicate the stretching (axial) direction.}} 
\label{figure2}
\end{figure}

Fully atomistic reactive molecular dynamics simulations were carried out to predict the tensile stress/strain behavior of CNTs, $\alpha$-GNTs, and $\gamma$-GNTs (Fig.~\ref{figure2}). We considered tubes with different diameters and chiralities. These 
simulations were performed using the LAMMPS (Large-scale Atomic/Molecular Massively Parallel Simulator) ~\cite{plimpton1995fast} code with the reactive Force Field (ReaxFF)~\cite{van2001reaxff}. There are many ReaxFF parameter sets, in the present work we used the parametrization described in \cite{PhysRevB.81.054103}. 
ReaxFF is a classical reactive potential suitable for studying fracture mechanics and breaking$/$formation. In contrast with more standard force fields, ReaxFF can describe breaking and bond formation. Its parameters are obtained from first-principles calculations of model structures and/or experimental data~\cite{van2001reaxff}. ReaxFF has been successfully used in the study of mechanical properties of nanostructured systems similar to those studied here~\cite{botari2014mechanical,cranford2011mechanical,nair2011minimal,de2016torsional,de2016mechanical,BIZAO2017431}.

Before starting the tube stretch (tensile) processes, in order to eliminate any residual stress present on the structures, we carried out an energy minimization followed by a NPT thermalization was carried. Then, the tensile tests were performed by stretching the nanotubes until fracture within NVT ensemble at room temperature (300 K). A chain of three Nos\'e-Hoover thermostats was used to control initial oscillations of the temperature~\cite{evans1985nose}. We used a constant engineering tensile strain rate $r=10^{-6}$/fs so that the nanotube length $L$ evolves as $L(t) = L_0(1+rt)$, where $L_0$ is the initial nanotube length. This strain rate is small enough to provide enough time to tube structural stabilization/reconstruction.

We considered 8 nanotubes of each type (CNT, $\alpha$-GNT, and $\gamma$-GNT), equally distributed between armchair and zigzag geometries. We also calculated the mechanical properties of CNTs with similar geometrical characteristics of the studied GNTs for comparison purposes. The selected CNTs $(n,0)$, and $(n,n)$ cases were $n=11,14,25,50$. 
In order to have GNTs with similar diameters, we chose $(n,0)$ and $(n,n)$ with $n=4,5,9,18$, for both $\alpha$-GNTs and $\gamma$-GNTs (Table~\ref{tubes}).

\begin{table}
\centering
\caption{Structural supercell geometrical information of the studied nanotubes.}
\vspace{0.5cm}
\label{tubes}
\begin{tabular}{cccc}
\hline 
CNT & Atoms & Diameter(\AA) & Length(\AA)\\
\hline
\hline
(11,0) & 352  & 8.61  & 34.08 \\
(14,0) & 448  & 10.96  & 34.08 \\
(25,0) & 800  & 19.57  & 34.08 \\
(50,0) & 1600  & 39.14  & 34.08 \\
(11,11) & 616  & 14.92  & 34.43 \\
(14,14) & 784  & 18.98  & 34.43 \\
(25,25) & 1400  & 33.90  & 34.43 \\
(50,50) & 2800  & 67.80  & 34.43 \\
\hline \\
\hline
$\alpha$-GNT & Atoms & Diameter(\AA) & Length(\AA)\\
\hline
\hline
(4,0) & 192  & 8.82  & 36.00 \\
(5,0) & 240  & 11.00  & 36.00 \\
(9,0) & 432  & 19.85  & 36.00 \\
(18,0) & 864  & 39.70  & 36.00 \\
(4,4) & 320  & 15.28  & 34.64 \\
(5,5) & 400  & 19.10  & 34.64 \\
(9,9) & 720  & 34.38  & 34.64 \\
(18,18) & 1440  & 68.75  & 34.64 \\
\hline \\
\hline
$\gamma$-GNT & Atoms & Diameter(\AA) & Length(\AA)\\
\hline
\hline
(4,0) & 288  & 8.71  & 35.54 \\
(5,0) & 360  & 10.89  & 35.54 \\
(9,0) & 648  & 19.60  & 35.54 \\
(18,0) & 1296  & 39.19  & 35.54 \\
(4,4) & 480  & 15.10  & 34.20 \\
(5,5) & 600  & 18.86  & 34.20 \\
(9,9) & 1080  & 33.94  & 34.20 \\
(18,18) & 2160  & 67.89  & 34.20 \\
\hline
\end{tabular}
\end{table}

During the stretching process we calculated the virial stress along the stretching direction $z$ ($\sigma_{z}$), as defined by~\cite{subramaniyan2008continuum,garcia2010bioinspired}:
\begin{eqnarray}
  \displaystyle \sigma_{z}=\sigma_{zz}=\displaystyle V^{-1}\sum_{k=1}^{N}(m_{k}v_{k_{z}}v_{k_{z}}+r_{k_{z}}f_{k_{z}}),
\end{eqnarray}
where $V=Ah = L_0\pi d_t h$ is the volume (considering a hollow cylinder) of the nanotube with $d_t$ being its diameter, $N$ the number of atoms, $m_{k}$ the mass, $v$ the velocity, $r$ the position, $f$ the force per atom, respectively. We adopted the standard graphene thickness  value of $h=3.34$ {\AA} (graphite interlayer distance) in all our calculations.

From the linear regime of each stress-strain curve we obtained the Young's modulus values, which are defined as:
\begin{eqnarray}
Y=\frac{\sigma_{z}}{\varepsilon_{z}},
\end{eqnarray}
where $\varepsilon_{z} = (L-L_0)/L_0$ is the strain along the $z$-direction (nanotube periodic axis) and $\sigma_{z}$ is the tensor component of the virial stress along the $z$-direction. We also calculated the normalized Young's modulus values ($Y_\rho=Y/\rho$) accordingly to the density of each structure ($\rho=m/(L_0\pi d_t h)$).

The spatial distribution of the stress during the stretching was calculated using the von Mises stress $\sigma_{v}^k$, defined as:
\begin{equation}
  \resizebox{.70\hsize}{!}{$\sigma_{v}^k=\sqrt{\frac{\left(\sigma_{11}^k-
          \sigma_{22}^k\right)^2+\left(\sigma_{22}^k-\sigma_{33}^k\right)^2+
        \left(\sigma_{11}^k-\sigma_{33}^k\right)^2+6\left((\sigma_{12}^k)^2+
          (\sigma_{23}^k)^2+(\sigma_{31}^k)^2\right)}{2}}$}.
\end{equation}


\subsection{DFT calculations}

In order to test the reliability of the MD results and also to validate them, we also carried out a systematic study of $\alpha$- and $\gamma$-GNTs under uniaxial strain using DFT methods~\cite{hohenberg64,Kohn65}, as implemented in the SIESTA (Spanish Initiative for Electronic Simulations with Thousands of Atoms) code \cite{ordejon96,portal97}. The Kohn-Sham orbitals were expanded in a double-$\zeta$ basis set composed of numerical pseudoatomic orbitals of finite range enhanced with
polarization orbitals. A common atomic confinement energy shift of 0.02 Ry was used to
define the cutoff radii of the basis functions, while the fineness of the real space grid was determined
by a mesh cutoff of 400 Ry~\cite{anglada02}. For the exchange-correlation potential, we used the generalized gradient approximation (GGA)~\cite{perdew96}. The pseudopotentials were modelled within the norm-conserving Troullier-Martins~\cite{troullier91} scheme in the Kleinman-Bylander~\cite{kleinman82} factorized form. Brillouin-zone integrations were performed using a Monkhorst-Pack~\cite{monkhorst76} grid of 1$\times$1$\times$8 k-points. All geometries were fully optimized for each strain value until the maximum force component on any atom was less than 0.01 eV/\AA. For each strained structural geometry relaxation, the SCF convergence thresholds for electronic total energy were set at 10$^{-4}$ eV.

Periodic boundary conditions were imposed, with perpendicular lattice vectors $a_x$ and $a_y$ large enough ($\sim$40\AA) to simulate vacuum and avoid spurious interactions between periodic images.
Similar to MD methodology, each strain level is defined as $\varepsilon_{z}=(L-L_0)/L_0$, where $L_0$ and $L$ hold for the relaxed and strained nanotube length, respectively.
Again, the graphyne nanotube was treated as a rolled membrane with thickness $h$ equal to 3.34 \AA{} and area equals to $\pi d_t L_0$, where $d_t$ is the nanotube diameter.
In the DFT study, the axial stress component $\sigma_{z}$ is related to the strain component $\varepsilon_{z}$ with the static relation $\sigma_{z}=(1/V)(\partial U/\partial\varepsilon_{z})$, where $V=L\pi d_t h$ is the volume of the strained nanotube membrane.

\section{Results}
The obtained critical strain ($\varepsilon_{c}$) and the ultimate strength (US) MD values for the studied nanotubes are presented in Table~\ref{tabcritical}. The complete structural failure (fracture) of both zigzag-alligned CNTs and $\gamma$-GNTs occurred around similar $\varepsilon_{c}$. On the other hand, different $\varepsilon_{c}$ were observed for armchair CNTs and $\gamma$-GNTs. Especially, $\alpha$-GNTs showed the highest $\varepsilon_{c}$ values for both zigzag and armchair nanotubes. We attributed these differences to the large pore size (see Fig.~\ref{figure1}), notably for $\alpha$-GNTs, and the characteristic bonding between acetylene groups in GNTs. $\gamma$-GNTs have hexagonal rings bonded to each other by acetylene groups that are not present in CNTs. This type of arrangement on GNTs significantly affects their mechanical properties.

\begin{table}
\centering
\caption{Critical strain ($\varepsilon_c$) and ultimate strength (US) values for CNTs, $\alpha$- and $\gamma$-GNTs.}
\vspace{0.5cm}
\label{tabcritical}
\begin{tabular}{|ccc|ccc|ccc|}
\hline 
CNT & $\varepsilon_{c}$ & US (GPa) & $\alpha$-GNT&$\varepsilon_{c}$ & US (GPa) &  $\gamma$-GNT&$\varepsilon_{c}$ & US (GPa) \\
\hline                                       
\hline                                            
(11,0)  & 0.16    & 122  &  (4,0)  & 0.24 &  45  & (4,0)  &  0.14&   81   \\
(14,0)  &  0.14   & 111  &  (5,0)  &0.24 &  49    & (5,0)  &  0.12&  68    \\
(25,0)  &   0.14  & 118  &  (9,0)  & 0.25& 46    &  (9,0)  & 0.13&    79  \\
(50,0)  &   0.13  & 115  & (18,0)  & 0.25& 45    & (18,0)  & 0.14&    81  \\
\hline   
(11,11) &  0.18  &  166 &  (4,4)  & 0.18&  44  & (4,4) &   0.11&      47\\
(14,14) &  0.16  &  170 &  (5,5)  & 0.18& 44    & (5,5)    & 0.09&      48\\
(25,25) &  0.17  & 167  &  (9,9)  & 0.19& 45    & (9,9)    & 0.11&      48\\
(50,50) &  0.18  & 166  & (18,18) & 0.18& 44    & (18,18)  & 0.12&      49\\
\hline
\end{tabular}
\end{table}

\begin{landscape}
\begin{table}
\centering
\caption{Young's modulus ($Y$) and normalized Young's modulus ($Y_\rho$) values for CNTs, $\alpha$- and $\gamma$-GNTs.}
\vspace{0.5cm}
\label{tab4}
\begin{tabular}{|ccc|ccc|ccc|}
\hline 
CNT & $Y$(GPa) & $Y_\rho$ $(\frac{\textrm{GPa.m}^{3}}{\textrm{kg}})$  & $\alpha$-GNT  & $Y$(GPa) & $Y_\rho$ $(\frac{\textrm{GPa.m}^{3}}{\textrm{kg}})$ &  $\gamma$-GNT & $Y$(GPa)& $Y_\rho$ $(\frac{\textrm{GPa.m}^{3}}{\textrm{kg}})$  \\
\hline                                                      
\hline                          

(11,0) &  710   &  0.3185  &  (4,0)  & 80  & 0.0711   & (4,0)   &   581   &  0.3386\\
(14,0) &  811  &  0.3662  &  (5,0)  & 77  & 0.0680   & (5,0)   &   525   &  0.3061\\
(25,0) &  821  &  0.3699  &  (9,0)  & 75  & 0.0668   &  (9,0)  &   657   &  0.3831\\
(50,0) &  881  &  0.3956  & (18,0)  & 70  & 0.0621   & (18,0)  &   633   & 0.3684\\
\hline    
(11,11)&  953  &  0.4298  &  (4,4)  & 171  & 0.1518   & (4,4)   &   430   &  0.2530\\
(14,14)&  998  &  0.4553  &  (5,5)  & 157  & 0.1414   & (5,5)   &   465   &  0.2761\\
(25,25)&  995  &  0.4532  &  (9,9)  &  145 & 0.1285   & (9,9)   &   463   &  0.2742\\
(50,50)&  911  &  0.4137  & (18,18) &  149 & 0.1325   & (18,18) &   472   & 0.2842\\

\hline
\end{tabular}
\end{table}
 \end{landscape}

\begin{figure}
\centering
\centerline{\includegraphics[width=10cm]{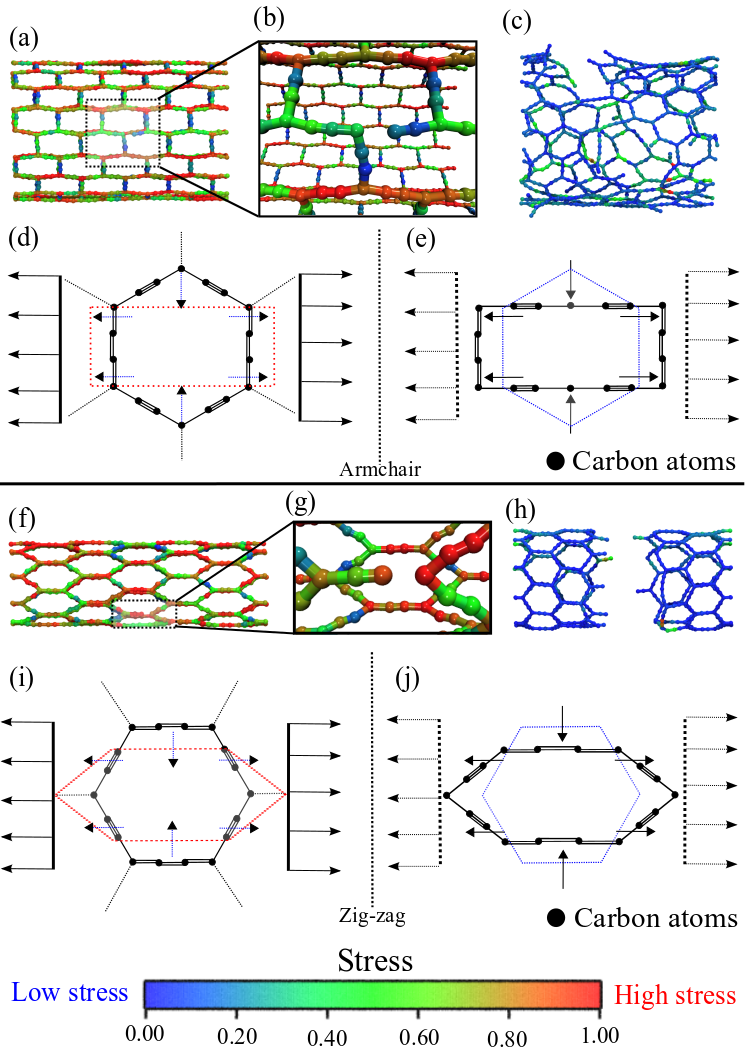}}
\caption{{\it Representative MD snapshots of a tensile stretch of the armchair $(4,4)$ (top) and zigzag $(4,0)$ (bottom) $\alpha-GNT$. (a,f) Lateral view of the strained nanotube colored accordingly to the von Mises stress values (low stress in blue and high stress in red). (b,g) Zoomed view of the starting of bond breaking. (c,h) MD snapshot of the nanotube just after fracture. Diagrams showing the ring dynamics of the armchair ((d) and (e)) and zigzag ((i) and (j)) $\alpha$-GNT can also be seen.}} 
\label{alfa}
\end{figure}

\begin{figure}
\centering
\centerline{\includegraphics[width=10cm]{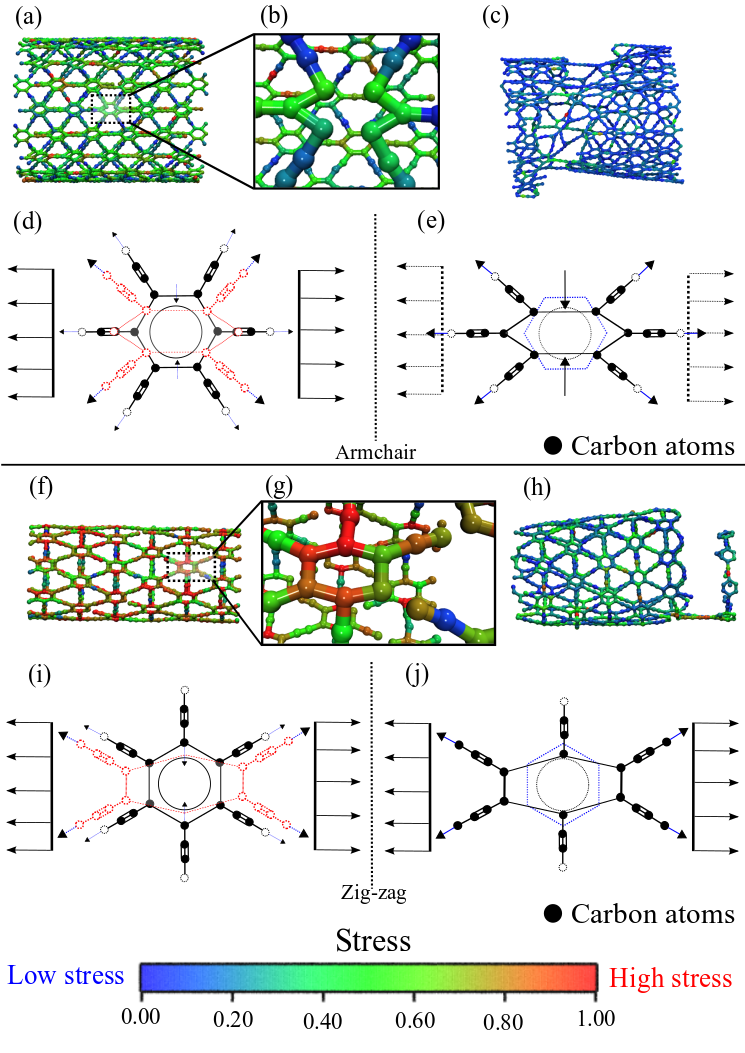}}
\caption{{\it Representative MD snapshots of a tensile stretch of the armchair $(4,4)$ (top) and zigzag $(4,0)$ (bottom) $\gamma-GNT$. (a,f) Lateral view of the strained nanotube colored accordingly to the von Mises stress values (low stress in blue and high stress in red). (b,g) Zoomed view of the starting of bond breaking. (c,h) MD snapshot of the nanotube just after fracture. Diagrams showing the ring dynamics of the armchair ((d) and (e)) and zigzag ((i) and (j)) $\gamma$-GNT can also be seen.}} 
\label{gama}
\end{figure}

\begin{figure}
\centering
\centerline{\includegraphics[width=10cm]{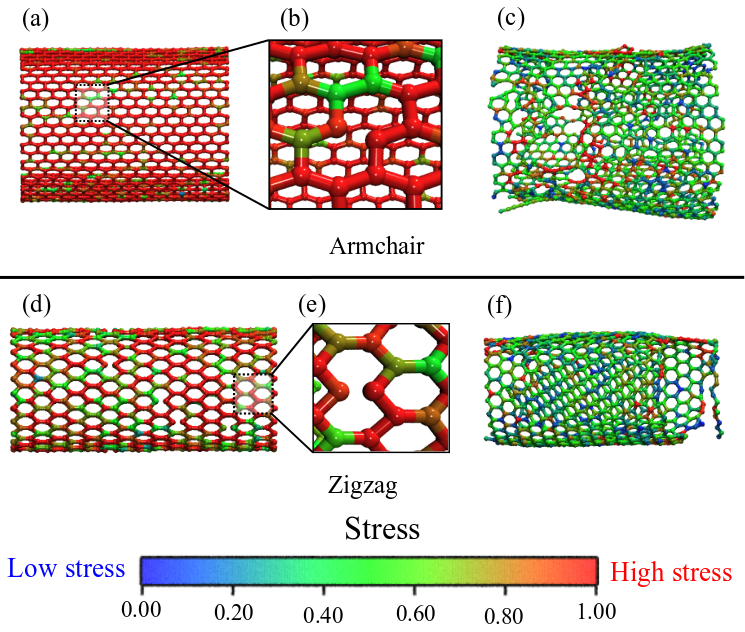}}
\caption{{\it Representative MD snapshots of a tensile stretch of conventional CNTs (armchair $(11,11)$ (top) and zigzag $(11,0)$ (bottom)). (a,d) Lateral view of the strained nanotube colored accordingly to the von Mises stress values (low stress in blue and high stress in red). (b,e) Zoomed view of the starting of bond breaking. (c,f) MD snapshot of the nanotube just after fracture.}}
\label{CNTtube}
\end{figure}

The GNTs and CNTs structural failure (fracture) processes can be better understood following the evolution of the von Mises stress distributions from the MD snapshots of the tensile stretch (Fig.~\ref{alfa}, Fig.~\ref{gama}, and Fig.~\ref{CNTtube}). From these Figures it is possible to observe high stress accumulation lines (in red) along the bonds parallel to the externally applied strain direction. These lines are composed of single and triple bonds in the armchair $\alpha$-GNTs and only by double bonds in the zigzag $\alpha$-GNTs (Fig.~\ref{alfa}). Fracture patterns of $\alpha$-GNTs indicate that bond breaking evolves initially from the single bonds for armchair nanotubes (highlighted rectangle of Fig.~\ref{alfa}(b)) until complete fracture (Fig.~\ref{alfa}(c)). As in the armchair case, the zigzag $\alpha$-GNT presents high stress accumulation along the chain of double bonds (as those are parallel to the nanotube main axis -- see Fig.~\ref{alfa}(f)), with bond breaking starting from these bonds (highlighted in Fig.~\ref{alfa}(g)). Figs.~\ref{alfa}(d),(e) show the covalent bonds of an armchair $\alpha$-GNT before and after applying strain along the nanotube main axis. The initial hexagonal shape of the pore changes to a more 
rectangular-like shape (red dashed rectangle in Fig.~\ref{alfa}(d)). This evolution patterns are consequences of the high tube flexibility, and it is responsible for the significant differences in the critical strain for armchair and zigzag $\alpha$-GNTs. Similar behavior was reported for graphyne membranes~\cite{zhao2013mechanical}. Because the pores of zigzag $\alpha$-GNTs (Fig.~\ref{alfa}(j)) are more flexible, they exhibit larger critical strain than armchair $\alpha$-GNTs (Fig.~\ref{alfa}(e)). The instants of the complete fracture are shown in Fig.~\ref{alfa}(h).

Similar results were also observed for armchair/zigzag $\gamma$-GNTs (Fig.~\ref{gama}) and CNTs (Fig.~\ref{CNTtube}). Figs. \ref{gama}(d),(e),(i),(j) show how the aromatic ring and the neighboring acetylene groups change during stretching. For the armchair $\gamma$-GNT, the six acetylene chains work as stress transmitter to the aromatic ring until the ring fracture. For the zigzag $\gamma$-GNT the fracture pattern is different with the fracture occurring on the single bonds of the acetylene groups. For CNTs (Fig.~\ref{CNTtube}), the stress is highly accumulated on the zigzag chains along the direction of the nanotube main axis, as in GNTs. The fracture starts from the bonds parallel and nearly parallel to the nanotube main axis for the zigzag and armchair CNTs, respectively. Because CNTs lack the acetylene chains, the structure is more rigid, the stress is accumulated directly on the hexagonal rings, the critical strains are smaller, and the ultimate strength value is larger. 

\begin{figure}
\centering
\centerline{\includegraphics[width=16cm]{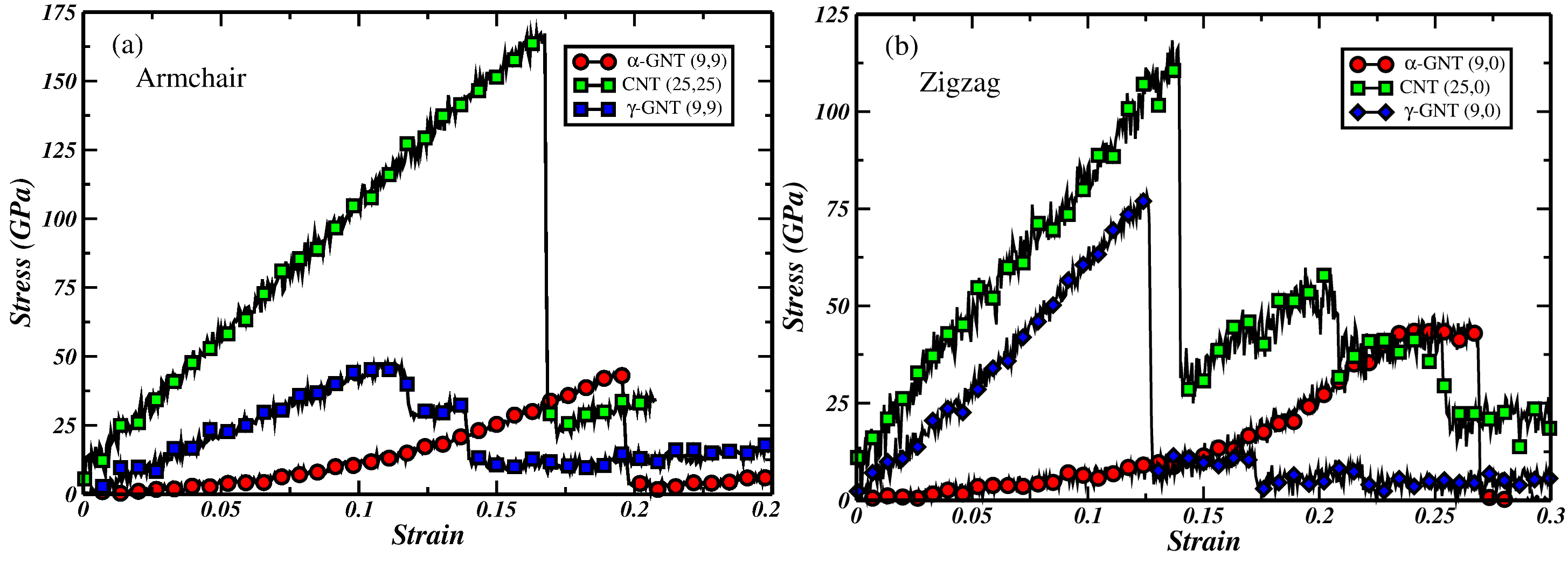}}
\caption{{\it Stress-strain curves obtained for CNTs ($(25,25)$ and $(25,0)$, green color), $\alpha$-GNTs ($(9,9)$ and $(9,0)$, red color), and $\gamma$-GNTs ($(9,9)$ and $(9,0)$, blue color) at 300 K}.} 
\label{stressStrain}
\label{Fig08}
\end{figure}

The stress-strain curves for $\alpha$-GNTs, $\gamma$-GNTs, and CNTs with similar diameters (Fig.~\ref{stressStrain}) are characterized by the existence of linear (elastic) and plastic regimes, where the bonds start to break until reaching a complete fracture, which is characterized by an abrupt stress drop. The Young's modulus values of the nanotubes studied here are presented in Table~\ref{tab4}. The obtained Young's modulus of the $(25,25)$ and $(25,0)$ CNTs were $995$ GPa and $821$ GPa, respectively, in good agreement with the average value $1020$ GPa of single-walled CNTs obtained by Krishnan \textit{et al.}~\cite{krishnan1998young}. As expected from the dynamics of the pore shape, $\gamma$-GNTs exhibit higher Young's modulus values when compared to $\alpha$-GNTs ones. The average Young's modulus values of the $(5,5)\ \gamma$-GNT calculated here was $465$ GPa, which is in agreement with a recent study, developed with the use of (AIREBO) potential, in which the obtained Young's modulus value was $466$ GPa \cite{li2018mechanical,Yang2012}.

The normalized Young's modulus $Y_\rho$ are presented in Table~\ref{tab4}. While the $\alpha$-GNTs values are small, the corresponding ones of $\gamma$-GNTs and CNTs are comparable, indicating that when the density is taken into account, in spite of their porosity, it is still possible to have graphyne structures with relative high  $Y_\rho$ values. This is clearly evidenced for the case of $(9,0)\ \gamma$-GNT, which possesses an even higher $Y_\rho$ in comparison to conventional CNTs.

\begin{figure}
\centering
\includegraphics[scale=0.75,angle={0}]{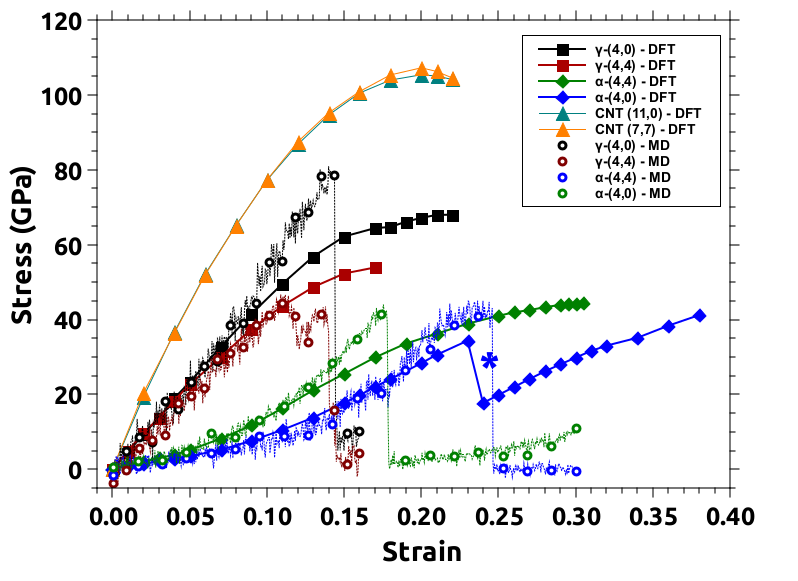}
\caption{Stress-strain curves obtained from DFT and MD calculations for conventional CNTs, $\alpha$-GNTs and $\gamma$-GNTs. The star symbol indicates a structural transformation at $\epsilon_{z} = 0.23$ found for the $(4,0)$ $\alpha$-GNT (see text for discussions). }
\label{dft1}
\end{figure}

In Fig. \ref{dft1} we present the stress-strain curves obtained from DFT calculations for some $\alpha$-GNTs, $\gamma$-GNTs, and CNTs superimposed with the same curves obtained from MD simulations. As we can see from the Figure, there is a good agreement between the methods, especially regarding the linear regime (Young's modulus vales) and ultimate strength values, although there is a tendency of larger values from DFT results.


The $\alpha$-GNTs are predicted to have the lowest Young's modulus value when compared to $\gamma$-GNTs and the corresponding CNTs. We also obtained similar fracture patterns from both MD and DFT methods. 
Interestingly, $\alpha$-GNTs become stiffer as they are stretched. The values of Young's modulus for $\alpha$-GNTs are almost twice from the unstrained configuration ($171$ GPa and $80$ GPa for (4,4) and (4,0), respectively) to strained configuration of $\varepsilon_z = 0.18-0.24$. The stress drop observed at $\varepsilon_z = 0.24$ for the (4,0) $\alpha$-GNTs is due to a structural transformation characterized by the formation of a triangle arrangement of carbon atoms due the large hexagonal stretching. Our calculations revealed that this transformation is diameter dependent. For instance, the transformation on the (5,0) and (9,0) $\alpha$-GNTs occurs at $\varepsilon_z=0.24$ and $\varepsilon_z=0.25$, respectively. Also, the large flexibility of $\alpha$-GNTs leads to critical strains up to 40\% of their original length and considerable toughness values, as can be seen in Table~\ref{tabtoughness}.

\begin{table}
\centering
\caption{Toughness ($\tau$) for different nanotubes calculated from DFT calculations}
\vspace{0.5cm}
\label{tabtoughness}
\begin{tabular}{|c|c|}
\hline 
Chirality & $\tau$ (GJ/m$^3$)  \\
\hline                                                      
\hline                                                      
$\alpha$-GNT (4,0)  & 7.68  \\
$\alpha$-GNT (4,4)  &  7.38  \\
$\gamma$-GNT (4,0)  &   9.63  \\
$\gamma$-GNT (4,4)  &   5.58  \\
CNT (7,7)  &   15.99  \\
CNT (11,0)  &  15.87  \\
CNT (20,0)  &  13.93  \\
CNT (12,12)  &  14.00 \\
\hline
\end{tabular}
\end{table}


\begin{figure}
\centering
\centerline{\includegraphics[width=12cm]{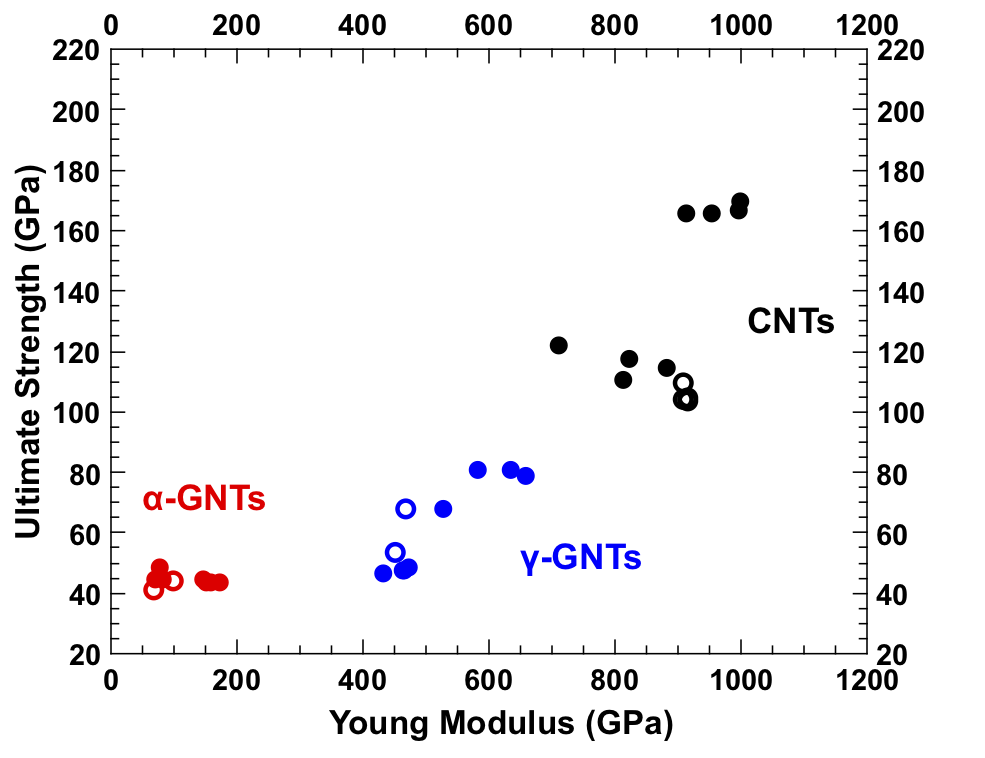}}
\caption{{\it Ultimate Strength (US) values as function of the Young's modulus values for different structures. CNTs, $\gamma$-GNTs and $\alpha$-GNTs are respectively in black, blue and red. Filled (open) symbols hold for MD (DFT) results.}} 
\label{Fig09}
\end{figure}

In Fig. \ref{Fig09} we plot the ultimate strength (US) as a function of the Young's modulus values for CNTs, $\gamma$-GNTs and $\alpha$-GNTs, for DFT and MD results. As we can see from this Figure, there is a good agreement between DFT and MD results as the structures occupy different niche values. With relation to these aspects, conventional CNTs perform better as they possess the highest Young's modulus and ultimate strength values, followed by $\gamma$-GNTs with intermediate values and lastly by $\alpha$-GNTs with lower values.

\section{Conclusions}

We investigated the structural and mechanical properties of graphyne  tubes (GNTs) of different diameters and chiralities, through fully atomistic reactive molecular dynamics and DFT calculations. We also considered conventional carbon nanotubes (CNTs), for comparison purposes.
Our results show that the complete structural failure (fracture) of
both zigzag-alligned CNTs and $\gamma$-GNTs occurred around similar critical strain values (($\varepsilon_{c}$)), but quite distinctly for armchair CNTs and $\gamma$-GNTs. In particular, $\alpha$-GNTs showed the highest ($\varepsilon_{c}$) values for both zigzag and armchair nanotubes. With relation to the fracture patterns, under stretch the 
stress accumulation occur along lines of covalent bonds parallel to the externally applied strain direction. These lines are composed of single and triple bonds in the armchair $\alpha$-GNTs 
and only by double bonds in the zigzag $\alpha$-GNTs. Similar results were obtained for $\gamma$-GNTs. The stress-strain curves for $\alpha$-GNTs, $\gamma$-GNTs, and CNTs with similar diameters
are characterized by the existence of linear (elastic) and plastic regimes, where the bonds start to break and propagate until reaching a complete fracture. But in contrast to CNTs, graphyne nanotubes exhibit significant diameter-dependent structural transitions after threshold ($\varepsilon_{c}$). With relation to the Young's modulus values, as expected CNTs exhibit larger values than graphyne nanotubes, but when these values are density normalized ($Y_\rho$), the graphyne and CNTs values are comparable, indicating that in spite of their porosity, it is still possible to have graphyne structures with relative high  $Y_\rho$ values. 
\section{Acknowledgements}

This work was supported by CAPES, CNPq, FAPESP, ERC and the Graphene FET Flagship.
J.M.S., R.A.B. and D.S.G. thank the Center for Computational Engineering and Sciences at Unicamp for financial support through the FAPESP/CEPID Grant $2013/08293-7$. J.M.S., A.G.S.F, A.L.A. and E.C.G. acknowledge support from PROCAD 2013/CAPES program. E.C.G. acknowledges support from CNPq (Process No. 307927/2017-2, and Process No. 429785/2018-6). J.M.S., A.L.A. and E.C.G. thank the Laborat\'orio de Simula\c c\~ao Computacional Caju\'ina (LSCC) at Universidade Federal do Piau\'i for computational support. V.R.C. acknowledges the financial support of FAPESP (Grant $16/01736-9$). A.L.A and V.P.S.F. acknowledges CENAPAD-SP for computer time and the Brazilian agencies CNPq (Process No. 427175/20160) for financial support. R.A.B. is supported by Fondazione Caritro under “Self-Cleaning Glasses” no. 2016.0278. NMP gratefully acknowledges the support of the grants by the European Commission Graphene Flagship Core 2 n. 785219 (WP14 ``Composites'') and FET Proactive ``Neurofibres'' n. 732344 as well as of the grant by MIUR ``Departments of Excellence'' grant L. 232/2016 and ARS01-01384-PROSCAN. 


\section{References}

\bibliographystyle{elsarticle-num}
\bibliography{Bibliografias}

\begin{thebibliography}{10}
\expandafter\ifx\csname url\endcsname\relax
  \def\url#1{\texttt{#1}}\fi
\expandafter\ifx\csname urlprefix\endcsname\relax\def\urlprefix{URL }\fi
\expandafter\ifx\csname href\endcsname\relax
  \def\href#1#2{#2} \def\path#1{#1}\fi

\bibitem{Novoselov666}
K.~S. Novoselov, A.~K. Geim, S.~V. Morozov, D.~Jiang, Y.~Zhang, S.~V. Dubonos,
  I.~V. Grigorieva, A.~A. Firsov,
  \href{http://science.sciencemag.org/content/306/5696/666}{{Electric Field
  Effect in Atomically Thin Carbon Films}}, Science 306~(5696) (2004) 666--669.
\newblock \href {http://dx.doi.org/10.1126/science.1102896}
  {\path{doi:10.1126/science.1102896}}.
\newline\urlprefix\url{http://science.sciencemag.org/content/306/5696/666}

\bibitem{thomas2008graphitic}
A.~Thomas, A.~Fischer, F.~Goettmann, M.~Antonietti, J.-O. M{\"u}ller,
  R.~Schl{\"o}gl, J.~M. Carlsson, Graphitic carbon nitride materials: variation
  of structure and morphology and their use as metal-free catalysts, Journal of
  Materials Chemistry 18~(41) (2008) 4893--4908.

\bibitem{zhang2015penta}
S.~Zhang, J.~Zhou, Q.~Wang, X.~Chen, Y.~Kawazoe, P.~Jena, Penta-graphene: A new
  carbon allotrope, Proceedings of the National Academy of Sciences 112~(8)
  (2015) 2372--2377.

\bibitem{wang2015phagraphene}
Z.~Wang, X.-F. Zhou, X.~Zhang, Q.~Zhu, H.~Dong, M.~Zhao, A.~R. Oganov,
  Phagraphene: A low-energy graphene allotrope composed of 5--6--7 carbon rings
  with distorted dirac cones, Nano letters 15~(9) (2015) 6182--6186.

\bibitem{baughman1987structure}
R.~Baughman, H.~Eckhardt, M.~Kertesz, Structure-property predictions for new
  planar forms of carbon: Layered phases containing sp2 and sp atoms, The
  Journal of chemical physics 87~(11) (1987) 6687--6699.

\bibitem{srinivasu2012graphyne}
K.~Srinivasu, S.~K. Ghosh, Graphyne and graphdiyne: promising materials for
  nanoelectronics and energy storage applications, The Journal of Physical
  Chemistry C 116~(9) (2012) 5951--5956.

\bibitem{hwang2013multilayer}
H.~J. Hwang, J.~Koo, M.~Park, N.~Park, Y.~Kwon, H.~Lee, Multilayer graphynes
  for lithium ion battery anode, The Journal of Physical Chemistry C 117~(14)
  (2013) 6919--6923.

\bibitem{lin2013mechanics}
S.~Lin, M.~J. Buehler, Mechanics and molecular filtration performance of
  graphyne nanoweb membranes for selective water purification, Nanoscale 5~(23)
  (2013) 11801--11807.

\bibitem{kou2014graphyne}
J.~Kou, X.~Zhou, H.~Lu, F.~Wu, J.~Fan, Graphyne as the membrane for water
  desalination, Nanoscale 6~(3) (2014) 1865--1870.

\bibitem{haley2008synthesis}
M.~M. Haley, Synthesis and properties of annulenic subunits of graphyne and
  graphdiyne nanoarchitectures, Pure and Applied Chemistry 80~(3) (2008)
  519--532.

\bibitem{cranford2011mechanical}
S.~W. Cranford, M.~J. Buehler, Mechanical properties of graphyne, Carbon
  49~(13) (2011) 4111--4121.

\bibitem{zhang2012mechanical}
Y.~Zhang, Q.~Pei, C.~Wang, Mechanical properties of graphynes under tension: a
  molecular dynamics study, Applied Physics Letters 101~(8) (2012) 081909.

\bibitem{peng2012mechanical}
Q.~Peng, W.~Ji, S.~De, Mechanical properties of graphyne monolayers: a
  first-principles study, Physical Chemistry Chemical Physics 14~(38) (2012)
  13385--13391.

\bibitem{kim2012graphyne}
B.~G. Kim, H.~J. Choi, Graphyne: Hexagonal network of carbon with versatile
  dirac cones, Physical Review B 86~(11) (2012) 115435.

\bibitem{de1995carbon}
W.~A. De~Heer, A.~Chatelain, D.~Ugarte, A carbon nanotube field-emission
  electron source, Science 270~(5239) (1995) 1179--1180.

\bibitem{harrison2007carbon}
B.~S. Harrison, A.~Atala, Carbon nanotube applications for tissue engineering,
  Biomaterials 28~(2) (2007) 344--353.

\bibitem{baughman1999carbon}
R.~H. Baughman, C.~Cui, A.~A. Zakhidov, Z.~Iqbal, J.~N. Barisci, G.~M. Spinks,
  G.~G. Wallace, A.~Mazzoldi, D.~De~Rossi, A.~G. Rinzler, et~al., Carbon
  nanotube actuators, Science 284~(5418) (1999) 1340--1344.

\bibitem{lima2012electrically}
M.~D. Lima, N.~Li, M.~J. De~Andrade, S.~Fang, J.~Oh, G.~M. Spinks, M.~E.
  Kozlov, C.~S. Haines, D.~Suh, J.~Foroughi, et~al., Electrically, chemically,
  and photonically powered torsional and tensile actuation of hybrid carbon
  nanotube yarn muscles, Science 338~(6109) (2012) 928--932.

\bibitem{iijima1991helical}
S.~Iijima, et~al., Helical microtubules of graphitic carbon, nature 354~(6348)
  (1991) 56--58.

\bibitem{dresselhaus1995physics}
M.~Dresselhaus, G.~Dresselhaus, R.~Saito, Physics of carbon nanotubes, Carbon
  33~(7) (1995) 883--891.

\bibitem{saito1998physical}
R.~Saito, G.~Dresselhaus, M.~S. Dresselhaus, et~al., Physical properties of
  carbon nanotubes, Vol.~35, World Scientific, 1998.

\bibitem{coluci2003families}
V.~Coluci, S.~Braga, S.~Legoas, D.~Galvao, R.~Baughman, Families of carbon
  nanotubes: Graphyne-based nanotubes, Physical Review B 68~(3) (2003) 035430.

\bibitem{coluci2004new}
V.~Coluci, S.~Braga, S.~Legoas, D.~Galvao, R.~Baughman, New families of carbon
  nanotubes based on graphyne motifs, Nanotechnology 15~(4) (2004) S142.

\bibitem{coluci2004theoretical}
V.~Coluci, D.~Galvao, R.~Baughman, Theoretical investigation of
  electromechanical effects for graphyne carbon nanotubes, The Journal of
  chemical physics 121~(7) (2004) 3228--3237.

\bibitem{li2018mechanical}
M.~Li, Y.~Zhang, Y.~Jiang, Y.~Zhang, Y.~Wang, H.~Zhou, Mechanical properties of
  $\gamma$-graphyne nanotubes, RSC Advances 8~(28) (2018) 15659--15666.

\bibitem{Reihani2018}
R.~Bizao, T.~Botari, E.~Perim, N.~M. Pugno, D.~Galvao,
  \href{http://www.sciencedirect.com/science/article/pii/S0008622317303743}{Graphyne
  nanotubes: Materials with ultralow phonon mean free path and strong optical
  phonon scattering for thermoelectric applications}, The Journal of Physical
  Chemistry C 122 (2018) 22688 -- 22698.
\newblock \href
  {http://dx.doi.org/https://doi.org/10.1016/j.carbon.2017.04.018}
  {\path{doi:https://doi.org/10.1016/j.carbon.2017.04.018}}.
\newline\urlprefix\url{http://www.sciencedirect.com/science/article/pii/S0008622317303743}

\bibitem{Zhao2015}
H.~Zhao, D.~Wei, H.~Shi, X.~Zhou, Thermal conductivities of graphyne nanotubes
  from atomistic simulations, Computational Materials Science 106~(1) (2010)
  69--75.

\bibitem{Yang2012}
Mechanical properties of graphyne and its family - a molecular dynamics
  investigation.

\bibitem{de2016torsional}
J.~M. de~Sousa, G.~Brunetto, V.~R. Coluci, D.~S. Galvao, Torsional
  “superplasticity” of graphyne nanotubes, Carbon 96 (2016) 14--19.

\bibitem{plimpton1995fast}
S.~Plimpton, Fast parallel algorithms for short-range molecular dynamics,
  Journal of computational physics 117~(1) (1995) 1--19.

\bibitem{van2001reaxff}
A.~C. Van~Duin, S.~Dasgupta, F.~Lorant, W.~A. Goddard, Reaxff: a reactive force
  field for hydrocarbons, The Journal of Physical Chemistry A 105~(41) (2001)
  9396--9409.

\bibitem{PhysRevB.81.054103}
T.~R. Mattsson, J.~M.~D. Lane, K.~R. Cochrane, M.~P. Desjarlais, A.~P.
  Thompson, F.~Pierce, G.~S. Grest,
  \href{https://link.aps.org/doi/10.1103/PhysRevB.81.054103}{First-principles
  and classical molecular dynamics simulation of shocked polymers}, Phys. Rev.
  B 81 (2010) 054103.
\newblock \href {http://dx.doi.org/10.1103/PhysRevB.81.054103}
  {\path{doi:10.1103/PhysRevB.81.054103}}.
\newline\urlprefix\url{https://link.aps.org/doi/10.1103/PhysRevB.81.054103}

\bibitem{botari2014mechanical}
T.~Botari, E.~Perim, P.~Autreto, A.~van Duin, R.~Paupitz, D.~Galvao, Mechanical
  properties and fracture dynamics of silicene membranes, Physical Chemistry
  Chemical Physics 16~(36) (2014) 19417--19423.

\bibitem{nair2011minimal}
A.~Nair, S.~Cranford, M.~Buehler, The minimal nanowire: mechanical properties
  of carbyne, EPL (Europhysics Letters) 95~(1) (2011) 16002.

\bibitem{de2016mechanical}
J.~M. de~Sousa, T.~Botari, E.~Perim, R.~Bizao, D.~S. Galvao, Mechanical and
  structural properties of graphene-like carbon nitride sheets, RSC Advances 6
  (2016) 76915--76921.

\bibitem{BIZAO2017431}
R.~Bizao, T.~Botari, E.~Perim, N.~M. Pugno, D.~Galvao,
  \href{http://www.sciencedirect.com/science/article/pii/S0008622317303743}{Mechanical
  properties and fracture patterns of graphene (graphitic) nanowiggles}, Carbon
  119 (2017) 431 -- 437.
\newblock \href
  {http://dx.doi.org/https://doi.org/10.1016/j.carbon.2017.04.018}
  {\path{doi:https://doi.org/10.1016/j.carbon.2017.04.018}}.
\newline\urlprefix\url{http://www.sciencedirect.com/science/article/pii/S0008622317303743}

\bibitem{evans1985nose}
D.~J. Evans, B.~L. Holian, The nose--hoover thermostat, The Journal of chemical
  physics 83~(8) (1985) 4069--4074.

\bibitem{subramaniyan2008continuum}
A.~K. Subramaniyan, C.~Sun, Continuum interpretation of virial stress in
  molecular simulations, International Journal of Solids and Structures 45~(14)
  (2008) 4340--4346.

\bibitem{garcia2010bioinspired}
A.~P. Garcia, M.~J. Buehler, Bioinspired nanoporous silicon provides great
  toughness at great deformability, Computational Materials Science 48~(2)
  (2010) 303--309.

\bibitem{hohenberg64}
P.~Hohenberg, W.~Kohn,
  \href{http://link.aps.org/doi/10.1103/PhysRev.136.B864}{Inhomogeneous
  electron gas}, Phys. Rev. 136 (1964) B864--B871.
\newblock \href {http://dx.doi.org/10.1103/PhysRev.136.B864}
  {\path{doi:10.1103/PhysRev.136.B864}}.
\newline\urlprefix\url{http://link.aps.org/doi/10.1103/PhysRev.136.B864}

\bibitem{Kohn65}
W.~Kohn, L.~J. Sham,
  \href{http://link.aps.org/doi/10.1103/PhysRev.140.A1133}{Self-consistent
  equations including exchange and correlation effects}, Phys. Rev. 140 (1965)
  A1133--A1138.
\newblock \href {http://dx.doi.org/10.1103/PhysRev.140.A1133}
  {\path{doi:10.1103/PhysRev.140.A1133}}.
\newline\urlprefix\url{http://link.aps.org/doi/10.1103/PhysRev.140.A1133}

\bibitem{ordejon96}
P.~Ordej\'on, E.~Artacho, J.~M. Soler, Self-consistent order-n
  density-functional calculations for very large systems, Phys. Rev. B 53
  (1996) 10441.

\bibitem{portal97}
D.~S\'anchez-Portal, E.~Artacho, J.~M. Soler, Density-functional method for
  very large systems with {LCAO} basis sets, Int. J. Quantum Chem. 65 (1997)
  453.

\bibitem{anglada02}
E.~Anglada, J.~M.~Soler, J.~Junquera, E.~Artacho,
  \href{http://link.aps.org/doi/10.1103/PhysRevB.66.205101}{Systematic
  generation of finite-range atomic basis sets for linear-scaling
  calculations}, Phys. Rev. B 66 (2002) 205101.
\newblock \href {http://dx.doi.org/10.1103/PhysRevB.66.205101}
  {\path{doi:10.1103/PhysRevB.66.205101}}.
\newline\urlprefix\url{http://link.aps.org/doi/10.1103/PhysRevB.66.205101}

\bibitem{perdew96}
J.~P. Perdew, K.~Burke, M.~Ernzerhof,
  \href{https://link.aps.org/doi/10.1103/PhysRevLett.77.3865}{Generalized
  gradient approximation made simple}, Phys. Rev. Lett. 77 (1996) 3865--3868.
\newblock \href {http://dx.doi.org/10.1103/PhysRevLett.77.3865}
  {\path{doi:10.1103/PhysRevLett.77.3865}}.
\newline\urlprefix\url{https://link.aps.org/doi/10.1103/PhysRevLett.77.3865}

\bibitem{troullier91}
N.~Troullier, J.~L. Martins, Efficient pseudopotentials for plane-wave
  calculations, Phys. Rev. B 43 (1991) 1993.

\bibitem{kleinman82}
L.~Kleinman, D.~M. Bylander,
  \href{https://link.aps.org/doi/10.1103/PhysRevLett.48.1425}{Efficacious form
  for model pseudopotentials}, Phys. Rev. Lett. 48 (1982) 1425--1428.
\newblock \href {http://dx.doi.org/10.1103/PhysRevLett.48.1425}
  {\path{doi:10.1103/PhysRevLett.48.1425}}.
\newline\urlprefix\url{https://link.aps.org/doi/10.1103/PhysRevLett.48.1425}

\bibitem{monkhorst76}
H.~J. Monkhorst, J.~D. Pack,
  \href{https://link.aps.org/doi/10.1103/PhysRevB.13.5188}{Special points for
  brillouin-zone integrations}, Phys. Rev. B 13 (1976) 5188--5192.
\newblock \href {http://dx.doi.org/10.1103/PhysRevB.13.5188}
  {\path{doi:10.1103/PhysRevB.13.5188}}.
\newline\urlprefix\url{https://link.aps.org/doi/10.1103/PhysRevB.13.5188}

\bibitem{zhao2013mechanical}
J.~Zhao, N.~Wei, Z.~Fan, J.-W. Jiang, T.~Rabczuk, The mechanical properties of
  three types of carbon allotropes, Nanotechnology 24~(9) (2013) 095702.

\bibitem{krishnan1998young}
A.~Krishnan, E.~Dujardin, T.~Ebbesen, P.~Yianilos, M.~Treacy, Young’s modulus
  of single-walled nanotubes, Physical Review B 58~(20) (1998) 14013.

\end{thebibliography}
\end{document}